\documentclass[utf8]{beilstein}
\usepackage{color}

\newcommand*{\blue}{\textcolor{black}}
\nolinenumbers
\begin{document}

\title{Nanoscale optical and structural characterisation of silk}
\author{Meguya Ryu}
\affiliation{Tokyo Institute of Technology, Meguro-ku, Tokyo 152-8550, Japan}
\author[1]{Reo Honda}
\author{Adrian Cernescu}
\affiliation{Neaspec GmbH, Bunsenstrasse 5, 82152 Martinsried, Germany}
\author{Arturas Vailionis}
\affiliation{Stanford Nano Shared Facilities, Stanford University, Stanford, CA 94305}
\author{Armandas Bal\v{c}ytis}
\affiliation{Swinburne University of Technology, John st., Hawthorn, 3122 Vic, Australia}
\author{\\\protect Jitraporn Vongsvivut}
\affiliation{Infrared Microspectroscopy Beamline, Australian Synchrotron, Clayton, Victoria 3168, Australia }
\author{Jing-Liang Li}
\affiliation{Institute for Frontier Materials, Deakin University, Geelong, VIC 3220, Australia}
\author[4]{Denver P. Linklater}
\author{Elena P. Ivanova}
\affiliation{School of Science, RMIT University, Melbourne, VIC 3001, Australia}
\author{\\\protect Vygantas Mizeikis}
\affiliation{Research Institute of Electronics, Shizuoka University, Naka-ku, 3-5-3-1 Johoku, Hamamatsu, Shizuoka 4328561, Japan}
\author[5]{Mark J. Tobin}
\author[1]{Junko Morikawa}
\author*[4]{Saulius Juodkazis}{sjuodkazis@swin.edu.au}
\affiliation{Tokyo Tech World Research Hub Initiative (WRHI), School of Materials and Chemical Technology, Tokyo Institute of Technology, 2-12-1, Ookayama, Meguro-ku, Tokyo 152-8550, Japan}
\affiliation{Melbourne Center for Nanofabrication, Australian National Fabrication Facility, Clayton~3168, Melbourne, Australia}
\maketitle
\begin{abstract}
\background Nanoscale composition of silk defining its unique properties via a hierarchial structural anisotropy has to be analysed at the highest spatial resolution of tens-of-nanometers
corresponding to the size of fibrils made of $\beta$-sheets, which
 are the crystalline building blocks of silk.
\results Nanoscale optical and structural properties of silk have been
measured from 100-nm-thick longitudinal slices of silk fibers with
$\sim 10$~nm resolution, the highest so far. Optical sub-wavelength resolution in
hyperspectral mapping of absorbance and molecular orientation were
carried out for comparison at IR wavelengths 2-10~$\mu$m using synchrotron radiation. 
\conclusion Reliable distinction of transmission changes by only 1-2\% due to anisotropy of amide bands was obtained from nano-thin slices of silk.
\end{abstract}

\section{Introduction}

Recent advances in nanofabrication of electronic devices require cutting-edge analytical technologies to provide a reliable structural characterisation of materials at the nanoscale. Such technologies are particularly important to probe molecular properties of cross-sections smaller than 100~nm in all three dimensions, which is of rapidly growing interest in the field of nanotechnology. Electronic chip manufacturing is currently introducing the sub-10~nm
fabrication node (a half pitch of a grating pattern) in the development of 3D fin-gates of field transistors. Nanofabrication techniques are now approaching single-digit-nm resolution using electron emission~\cite{Ivo} and thermal probes~\cite{Paul,Holzner}.
Further control of surface nano-texturing, to achieve regularly patterned features with sub-100~nm resolution, is
currently under development for inherent material properties, such as controllable
surface wettability, anti-biofouling, anti-reflection, and
biocidal/bactericidal properties~\cite{17n245301,13nc2838}. For example, the motheye plastic films produced by roll-to-roll technology already replicate nano-pillars with 100~nm separation (MOSMITE
from Mitsubishi Chemicals. Ltd.).

The structural and optical properties of material are interrelated. By using a wide spectrum of electromagnetic waves from visible to THz, it is possible to gain insights into complex hierarchical structures of composite materials. For materials with strong structural anisotropy, defined by the molecular orientation and alignment of crystalline micro-volumes, it is important to characterise structure at the highest lateral and longitudinal resolutions\blue{~\cite{silksci,tiger}}. Anderson localisation of light and thermal cooling of silk at IR wavelengths was recently demonstrated to be related to the fibril sub-structure of silk, which was in the range of tens-of-nanometers~\cite{Local}. This defines the range of the spatial
resolution required for structural and
chemical analysis, which are typically carried out using X-ray and IR based techniques at larger scales.

Real and imaginary parts of the refractive index, $\tilde{n}= n + i\kappa$, together with the orientation dependency of the
birefringence $\Delta n$ and dichroism $\Delta\kappa$, define the optical response of materials. Reflectance $R$ is
proportional to the real part, while the absorbance $A$ corresponds to the
imaginary part of $\tilde{n}$. Recently, we demonstrated that the IR measurements of silk performed using three different methods, including \emph{(i)} a table-top Fourier
transform infrared (FTIR) transmission, \emph{(ii)} a
synchrotron-based attenuated total reflection (ATR) FTIR, and
\emph{(iii)} an atomic force microscopy (AFM) tip response to the
IR absorbed light (nano-IR~\cite{Dazzi}), produced comparable spectral features~\cite{17mre115028}. Whilst the first two modalities probe micron-sized volumes of silk, the AFM-based Nano-IR technique acquires structural information at the nanoscale (i.e. the area under the AFM tip \blue{from} volume \blue{with lateral cross-section of} $\sim 20$~nm). Differences in absorbance and spectral line-shapes of the characteristic silk bands are related to the
different sensitivity of $R$ and $A$ to the real and imaginary
parts of $\tilde{n}= n + i\kappa$. The absorbance measured from
the far-field transmission directly reflects
the imaginary part of the index $\kappa$, while those obtained in the ATR-FTIR mode are affected by the real part of the index \textit{via} the Snell's law~\cite{vib}. As a result, comparative measurements of the absorbance by different near- and far-field techniques are essentially required to understand differences in electric-field determination of the local light and its interaction with the sample~\cite{Huth}.  

\begin{figure}
\begin{center}
\includegraphics[width=15cm]{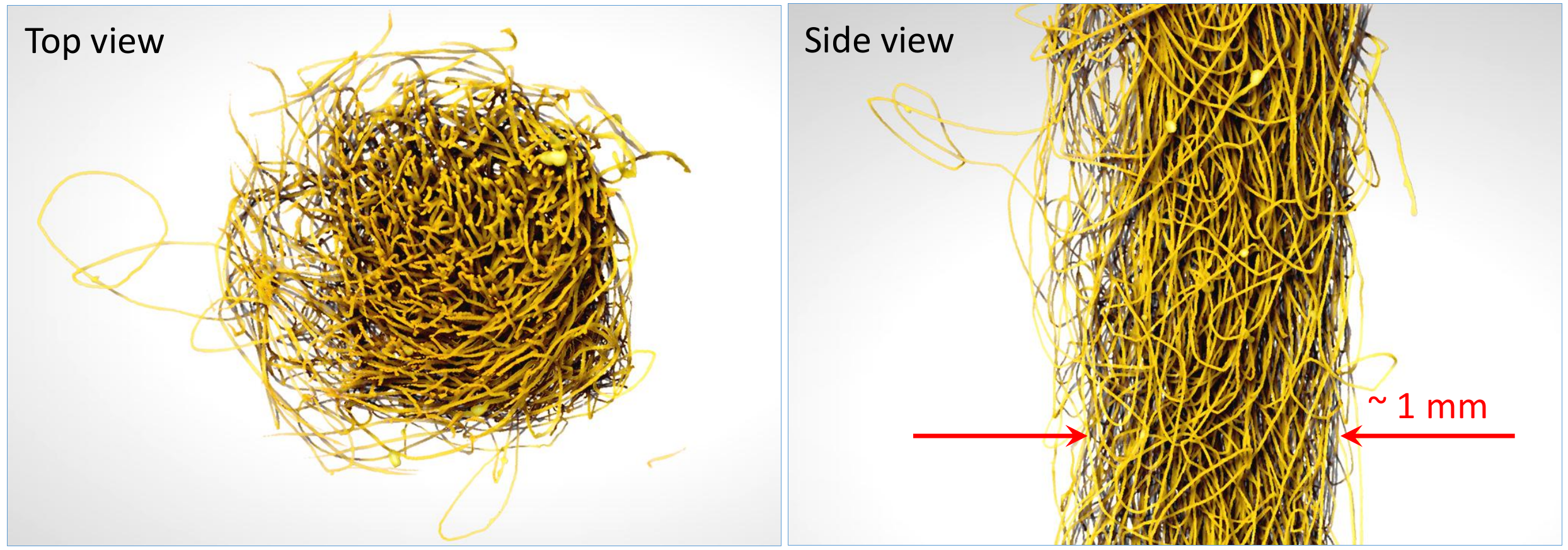}
\caption{X-ray tomographic images showing 3D rendered volume of white
silk  \emph{Bombyx mori} fibers at 3.15~$\mu$m pixel resolution.
The bundle of silk fibers was made of degummed single strand silk fibers. }
\label{f-bunch}
\end{center}
\end{figure}

\blue{Different modalities of sample preparation for nanoscale imaging include focused ion beam milling and micro-tome slicing. When thickness of samples, especially soft bio-materials, becomes close to 100~nm the cutting tool might cause a tear and cut induced strain under the surface. In turn, this can cause artefacts in determination of optical properties which are related to the mass density and its gradients. It is important to measure $n$ and $\kappa$ from decreasingly smaller volumes and to compare with data obtained from the bulk samples.}


Here, we used a near-field scattering method to probe $n, \kappa$ and to
determine spectral differences between the reflectance and absorbance
of silk fibers with $\sim 10$~nm resolution. Cross-sections of silk fibers
were prepared using an ultramicrotome. Silk was chosen due to its well known spectral properties
and its increasing applications as a biocompatible and biodegradable
material~\cite{Ling,Hotz}. \blue{Silk has uniaxial symmetry which can be examined from longitudinal microtome slices used in this study.}  Sub-wavelength resolutions in
hyperspectral IR mapping of absorbance and orientational
properties of the absorbing bands were reliably measured from
100-nm-thick slices of silk. Such a high resolution technique is essential in order to gain a better understanding of the fibril structure of silk~\cite{Local}.
\begin{figure}
\begin{center}
\includegraphics[width=14cm]{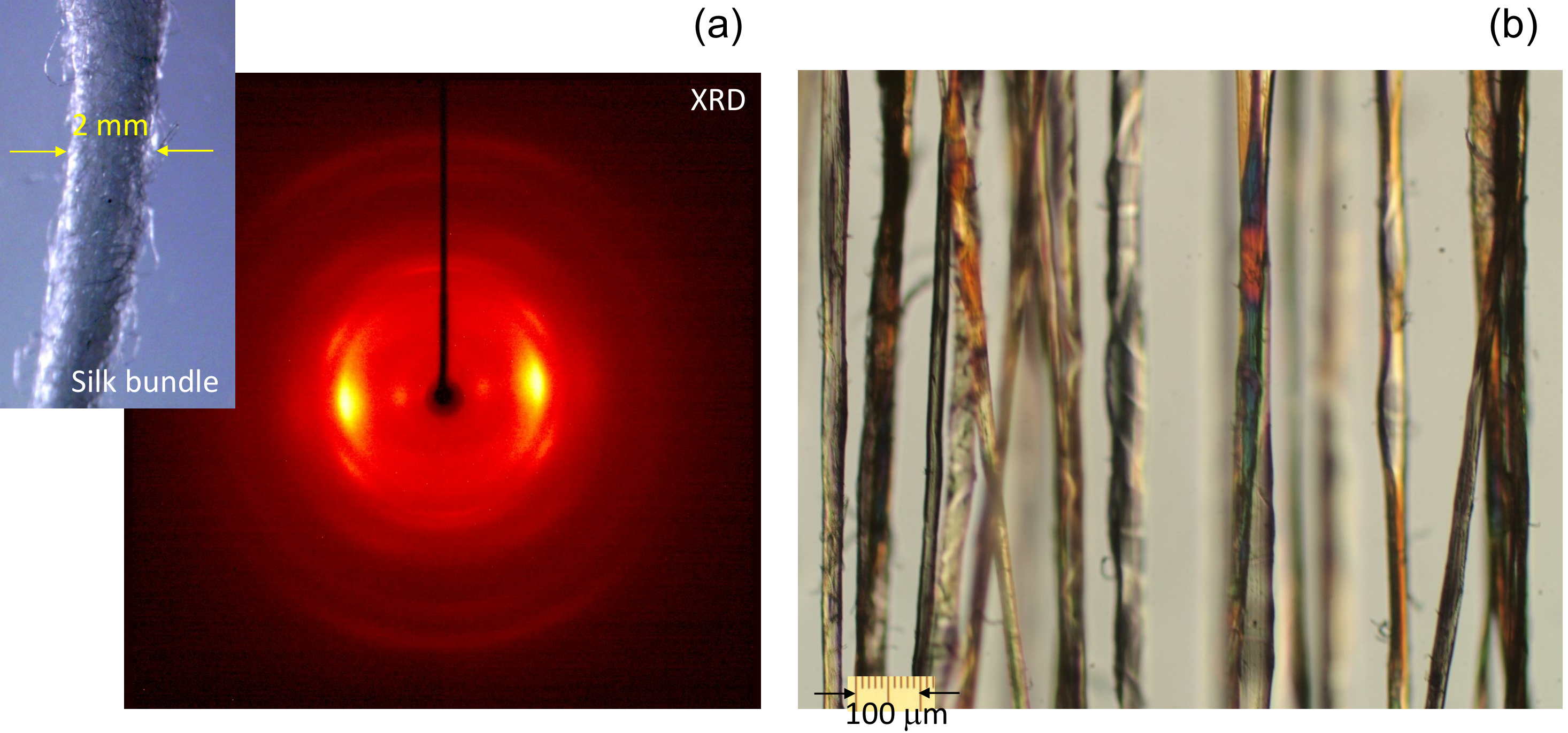}
\caption{(a) Wide-angle 2D x-ray diffraction of a bundle of white silk \emph{Bombyx mori} fibers. Inset shows an optical microscopic
image of \blue{a convolved} silk \blue{fiber} bundle. The silk bundle was made of degummed single
strand silk fibers. The long axis of the fibers was predominantly
vertical. (b) Optical image of white silk fibers through optically
aligned polariser-analyser (high transmission) setup under white
light illumination using a Nikon MPlan 10$^\times$ DIC objective lens
with numerical aperture $NA = 0.25$. } \label{f-silk}
\end{center}
\end{figure}

\section{Methods and samples}
\subsection{Silk slices}

Silk fibers were embedded in epoxy resin (Oken Ltd., Japan) and thin-sectioned by ultramicrotomy to achieve a sample thickness of  $\sim 100$~nm. Slices were then immobilised on IR-transparent non-birefringent substrates (CaF$_2$). 

\subsection{X-ray characterisation}

A 3D X-ray computed micro-tomography (micro-CT) was performed using a lab-based ZEISS Versa 520 X-ray Microscope at the Stanford Nano Shared Facilities, Stanford University. The scan settings were as follows: source voltage - 30~kV, pixel size – 3.15~$\mu$m, number of projections – 1600, exposure time – 10~s. The micro-CT dataset was reconstructed using ZEISS Scout-and-Scan Reconstructor software (Fig.~\ref{f-bunch}).

2D X-ray diffraction of \emph{Bombyx mori} silk was carried out on a Bruker D8 Venture single crystal diffractometer using a Cu-Ka micro-focus X-ray source with wavelength of $\lambda = 1.5418\AA$ (Fig.~\ref{f-silk}(a)).

\subsection{IR spectral measurements}


The sub-diffraction scattering scanning near-field optical microscopy s-SNOM (neaspec GmbH) uses a metalized atomic force microscopy (AFM) tip which maps the surface relief (topography) by
its basic AFM operation, and simultaneously, under external infrared illumination (broadband laser working by difference-frequency generation, Toptica), acts as a light-concentrating antenna such that the sample is probed with a nanofocused light field. The AFM tapping mode operation (ca. 60~nm amplitude) modulates the near-field interaction between the tip
and sample~\cite{Hillenbrand}. An asymmetric Michelson interferometer and a lock-in detection of the signal at higher harmonic of the tapping frequency (approximately 250~kHz) provides
background-free nano-IR spectra and images with maximum resolution imposed by the AFM tip size independent of the laser wavelength~\cite{Huth}.

The nano-FTIR spectra were recorded in ca. 100~s/spectrum with a spectral resolution of 10~cm$^{-1}$. Removal of the instrumental response function from the nano-FTIR spectra was done
by normalization of the measured spectra to a reference Si signal. Resulting nano-FTIR absorption and reflectivity spectra can be directly correlated with the standard far-field IR spectra~\cite{Govyadinov,Westermeier}.

\begin{figure}
\begin{center}
\includegraphics[width=18cm]{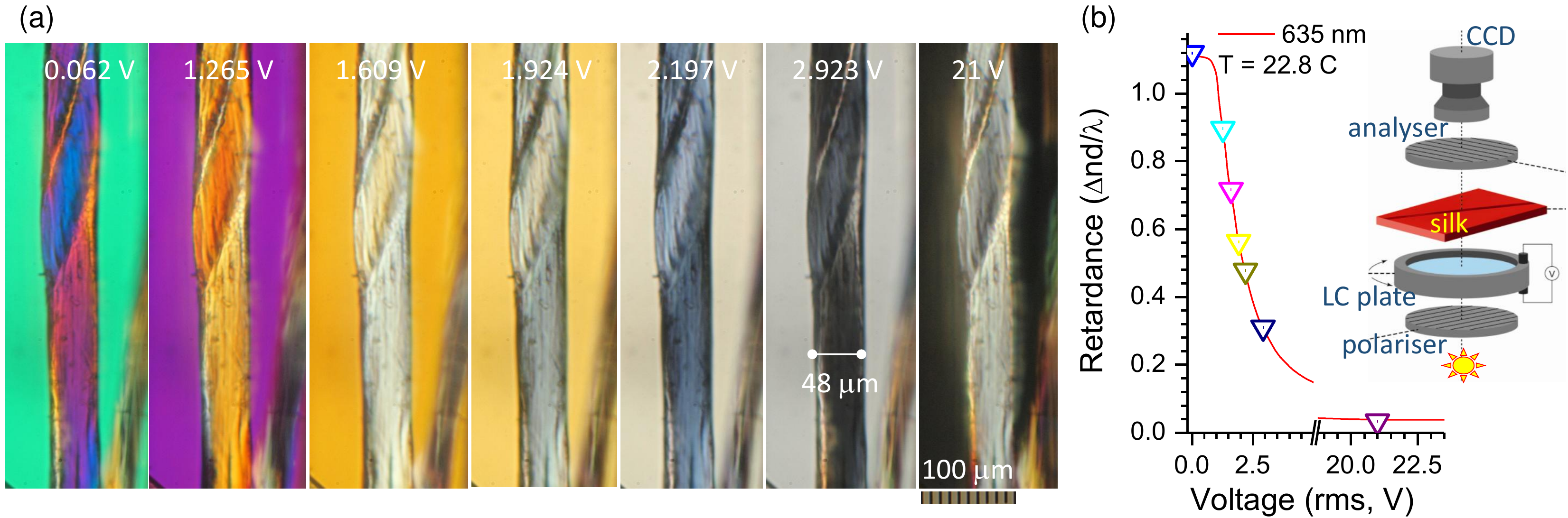}
\caption{(a) A series of optical images taken at different voltages of a
liquid crystal (LC) retarder (schematically shown in the inset of (b)) and a Nikon Optiphot-pol
microscope with LMPlanFL 20$^\times$ objective lens $NA = 0.4$. (b) Calibration curve of retardance as a function of voltage collected at 635~nm wavelength and 22.8$^\circ$C.}\label{f-retar}
\end{center}
\end{figure}

Hyperspectral imaging of the absorbance was measured on the IR Microspectroscopy (IRM) Beamline at Australian Synchrotron (Victoria, Australia). The
measurements were performed using a Bruker Hyperion 2000 FTIR microscope (Bruker Optik GmbH, Ettlingen, Germany) coupled to a Vertex  V80v  FT-IR spectrometer equipped with a liquid
nitrogen-cooled narrow-band mercury cadmium telluride (MCT) detector. Holographic ZnSe wire-grid polarisers (Edmund) were used to set polarisation at the IR spectral range of $\lambda = 4000 - 750$~cm$^{-1}$ (2.5 - 13.3~$\mu$m); the extinction of polarisers was $T^{max}/T^{min}\simeq$150 and transmittance $\sim 50$\%. The far-field transmission measurements were carried out with a numerical aperture $NA = 0.5$, $36^\times$ magnification Cassegrain objective lens at the corresponding resolution of $0.61\lambda/NA \simeq 4.1~\mu$m at the 3000~cm$^{-1}$ band ($\lambda =$3.33~$\mu$m). The absorbance or optical density $A = -lg(T)$ spectrum is defined by the absorption
coefficient $\alpha\equiv 4\pi\kappa/\lambda =
2\omega\kappa/c$~[cm$^{-1}$] for the transmitted light intensity $I_T = I_0e^{-\alpha d} = I_0\times 10^{-OD}$; where $d$ is thickness of sample, transmittance $T = I_T/I_0$, $OD$ is optical density,
$\omega$ is the cyclic frequency of light, and $c$ is speed. The reflectance for the normal incidence from air is defined as $R = [(n-1)^2 + \kappa^2]/[(n+1)^2 + \kappa^2$].

\begin{figure}
\begin{center}
\includegraphics[width=16cm]{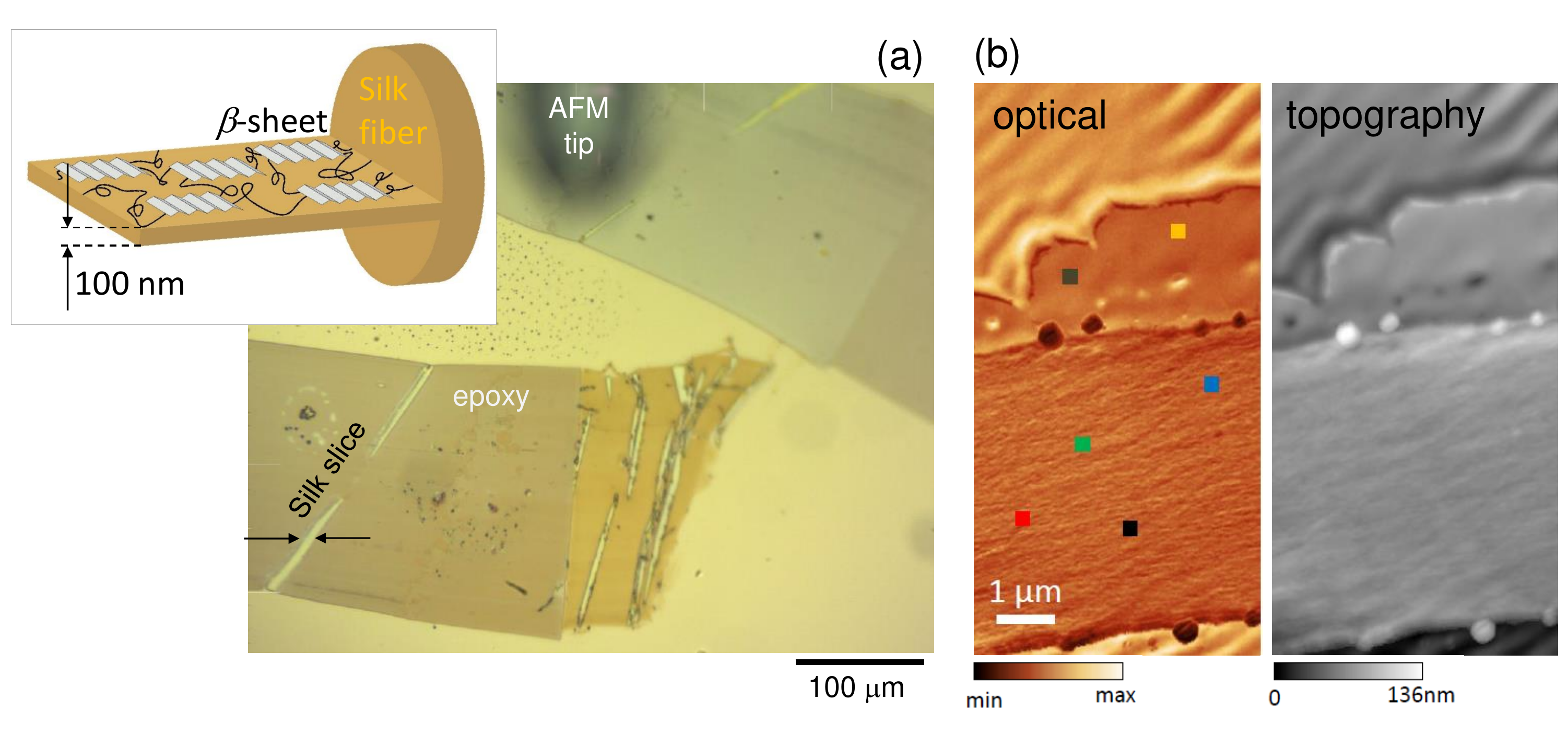}
\caption{(a) Far-field optical image of longitudinal
slices of white silk embedded in an epoxy sheet. \blue{The inset shows schematics of a lateral silk slice composed of $\beta$-sheets interconnected with $\alpha$-coils and amorphous segments.} (b) Optical and topographic images
of the silk slice shown in (a) measured with scattering near-field
microscopy (SNOM; neaspec). Markers in optical image indicate
locations where spectra were acquired.} \label{f-tome}
\end{center}
\end{figure}
\begin{figure}
\begin{center}
\includegraphics[width=16cm]{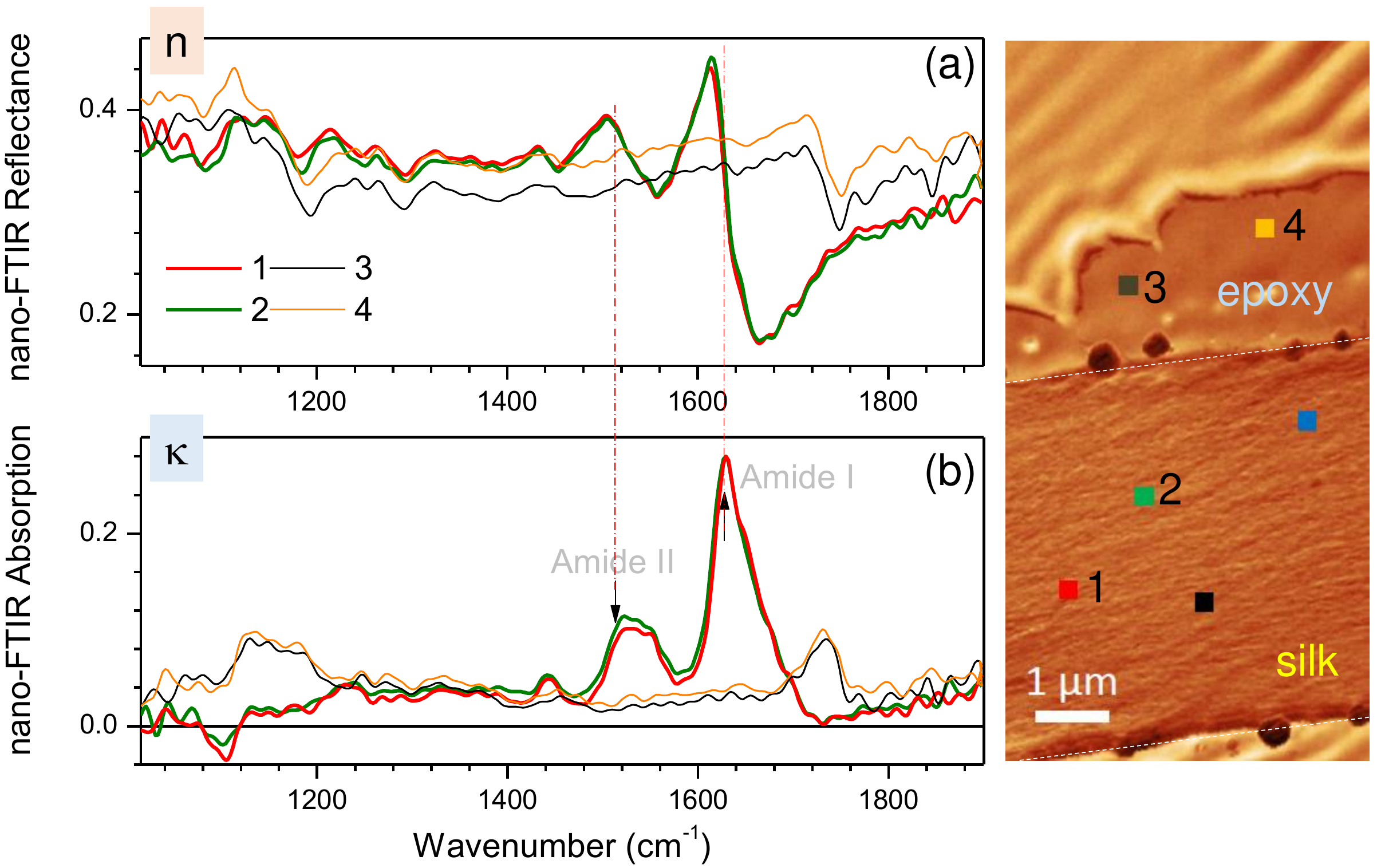}
\caption{Scattering near-field optical microscopy (SNOM)
measurements of the nano-FTIR
reflectance (a) and absorption (b) spectra
from selected points on silk and epoxy (shown in the right inset).} \label{f-spectra}
\end{center}
\end{figure}
\begin{figure}
\begin{center}
\includegraphics[width=18.5cm]{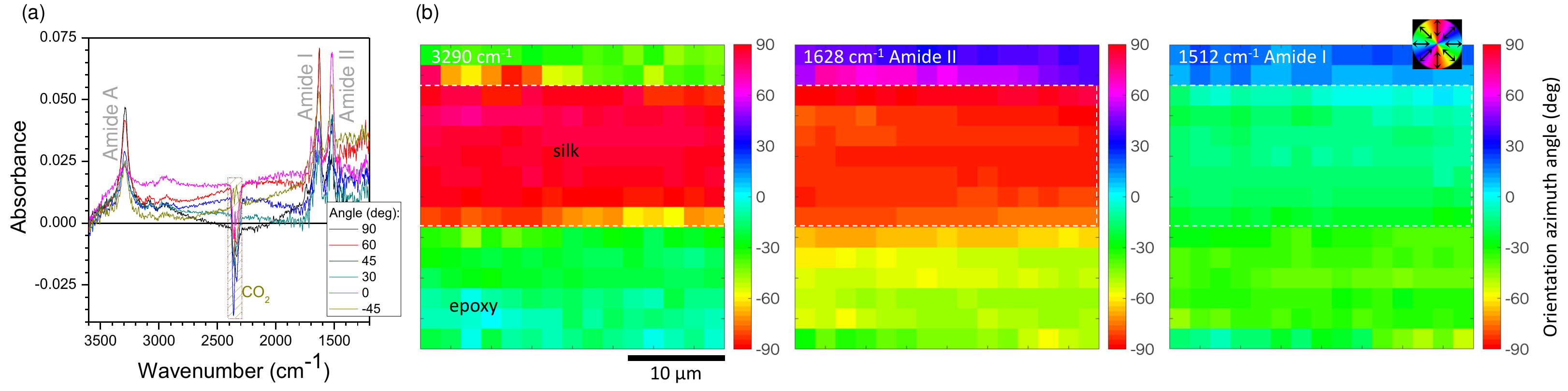}
\caption{(a) Single point absorbance spectra of thin-sectioned silk on BaF$_2$ collected at
different angles $\theta$ between the linear
polarisation and the fiber axis, using 2~$\mu$m pixel pitch, $15\times 15$ pixel points,
4.17~$\mu$m spatial resolution, and 4~cm$^{-1}$ spectral resolution. (b) Orientation color maps indicating that Amide A
(N-H) is oriented perpendicular to Amide I (C=O) and Amide II (C-N) bands. } \label{f-100}
\end{center}
\end{figure}

\section{Results and discussion}

X-ray diffraction is the method of choice to reveal the internal structure of complex materials and to detect crystalline regions. Figures~\ref{f-bunch} and \ref{f-silk}(a) show 3D reconstruction of the \emph{Bombyx mori} silk fibers bundled together and their X-ray diffraction (XRD) pattern, respectively. The period $d$ corresponds to the most pronounced peaks at the diffraction angle $2\theta$, given by the Bragg's condition $d =
\lambda/(2\sin\theta)$. The size $L$ of the nano-crystalline phase can be estimated from the Scherrer's equation $L =
K\lambda/(B(2\theta)\cos\theta$; where $K = 0.89$ for spherical crystals and $B(2\theta)$ is the peak's angular bandwidth at full width half maximum. The wide angle XRD pattern (Fig.~\ref{f-silk}(a)) is identical to that reported earlier~\cite{Drummy}. The most pronounced peak corresponds to the separation between the equatorial (200) planes $d_{(200)} =
4.69$~nm and crystal cross-section of $L\simeq 2.15$~nm, while for the meridional (002) planes $d_{(002)} = 3.46$~nm and crystal size
of $L\simeq 10.76$~nm~\cite{Drummy}. These are dimensions of the $\beta$-sheets, which are crystalline segments in the silk fiber. \blue{SNOM measurements are well suited to measure $n$ and $\kappa$ from areas of comparable dimensions.}

Silk is a strongly birefringent material, as revealed by cross-polarised optical imaging (Fig.~\ref{f-retar}). The images were taken following adjustment of the voltage of a liquid crystal
(LC) retarder, which was inserted with its slow-axis perpendicular to the orientation of the silk fiber (see inset in (b)). Using such a geometry, it is possible to compensate for the silk fiber birefringence, $\Delta n\equiv n_e - n_o > 0$, with a phase delay imparted by the LC retarder. When the phase delay through the LC retarder is equal to
the absolute value, but has an opposite sign through the silk fiber, the most dark (black) region is formed in the image at $\sim 2.9$~V (Fig.~\ref{f-retar}(a)). For the thickness of
fiber $d = 48~\mu$m and measured retardance, the birefringence $\Delta n \simeq 4\times 10^{-3}$. This is an order of magnitude estimate since LC retarder calibration is made at single wavelength, while the imaging is captured under white light illumination. The birefringence originates from the structure, which has a fiber-orientation-determined alignment down to molecular bonds and spans hierarchically over a wide range of wavelengths due to secondary ordering~\cite{17m356}. Previously, longitudinal $\sim 1~\mu$m-thick silk slices were measured in transmission mode using synchrotron IR radiation to characterise the molecular alignment of
the typical amide bands~\cite{17sr7419}, including amide II at 1512~cm$^{-1}$ (C-N), amide I ($\beta$-sheets) at 1628~cm$^{-1}$ (C=O), and amide A at 3290~cm$^{-1}$ (N-H). Perpendicular
orientation between C=O and C-N bonding was revealed at a high accuracy when longitudinal silk slices were prepared~\cite{17sr7419}. \blue{Longitudinal slices facilitated more precise measurements of molecular alignment since there were no averaging artefacts due to the curvature of silk fiber and different thickness across the fiber slice~\cite{recent}.}

Scattering SNOM was used to measure reflectance and absorbance spectra from nanoscale areas of a single silk slice. Lateral slices of
0.1~$\mu$m were prepared on a gold mirror (Fig.~\ref{f-tome}(a)).
Optical and topographic images were obtained, which confirmed the thickness of the silk slices to be $\sim 100$~nm (Fig.~\ref{f-tome}(b)). Spectra of
nano-FTIR reflectance and absorption from selected points were also measured (Fig.~\ref{f-spectra}) with a high reproducibility, showing a clear distinction between the silk and host-epoxy matrix. The nano-FTIR absorption is proportional to the
imaginary part of the scattering coefficient $\sigma_n(\omega) = s(\omega)e^{i\phi(\omega)}$, which relates the light scattered field $E_s(\omega)$, and the incident field $E_i(\omega)$ through the equation $E_s=\sigma_n E_i$; where $s(\omega)$ and $\phi{(\omega)}$ are the amplitude and phase of the back-scattered spectra~\cite{Huth}.
The reflectivity information is given by the real part of the scattering coefficient~\cite{Huth}. Using the asymmetric Michelson interferometer, the full complex function of the scattered optical signal could be recorded, therefore enabling the
simultaneous measurement of both nano-FTIR absorption and reflectivity spectra, shown in Fig.~\ref{f-spectra}.
 
The amide I and II bands were well reproduced in absorption spectra collected from four different single points; however, only spectra from two measurement points are displayed in
Fig.~\ref{f-spectra} for a better clarity of presentation. Nanoscale resolution is readily achievable for  SNOM measurements and is defined by the AFM tip, which has a diameter of $\sim 10$~nm. \blue{Around the center of the absorption peak, regions of normal dispersion with a higher refractive index at a higher photon energy (proportional to wavenumber) was observed. Spectral positioning of the  absorption peak and dispersion lineshapes corresponded to the expected Lorenzian behaviour of a damped oscillator. }

Next, direct absorbance and orientation mapping~\cite{Hikima}
through a 100-nm-thick silk slice was demonstrated using
synchrotron IR radiation (Fig.~\ref{f-100}). By measuring absorbance
at several azimuth angles, $\theta$, it was possible to determine
molecular alignment within the fibril structure. Here, we demonstrate the use of the technique on the
thinnest silk section \blue{of 100~nm}. The well aligned amide bands were \blue{measured} in transmission mode at wavelengths, which are much longer than the thickness of the silk slice ($d = 100$~nm) when compared to the wavenumber
1500~cm$^{-1}$ equivalent to 6.67~$\mu$m wavelength. The pitch between measurement points was $2~\mu$m and was approximately two times smaller than the focal spot ($4.1~\mu$m). This caused an uncertainty in orientation azimuth at the boundary of the silk fiber and the surrounding epoxy matrix.
However, the central part of the fiber shows a well-defined
orientation, while the epoxy region has a random
orientation. The absorbance from silk, which makes only $d/\lambda
\simeq 1.5\%$ of the probing wavelength, was reliably measured \blue{in transmission}. The retardance of silk, $d = 100$~nm, has $\Delta n = 4\times 10^{-3}$ birefringence at the non-absorbing visible-IR wavelengths. For example, the band at 3600~cm$^{-1}$ ($\lambda = 2.78~\mu$m) resulted in $\Delta
T = \sin^2(\pi\Delta n d/\lambda) = 2\times 10^{-5}\%$, which was
beyond the precision of measurements. Alternatively, the real part refractive index can be determined from the known reflectance $R$ and extinction $\kappa$ values following the equation $n = [(1+R)/(1-R)] + [4R/(1-R)^2-\kappa^2]^{1/2}$; however, $R$ was not measured in this experiment. 

Anisotropy in absorption is defined by the dichroism $\Delta"\equiv(A_{\parallel}-A_{\perp})/2 =
k(\kappa_{\parallel} - \kappa_{\perp})d$;  where $k =
2\pi/\lambda$ is the wavevector. It defines the losses in transmission $T$, at the maximum and minimum orientations of linear
polarisation $e^{-\Delta^{"}} = \sqrt{T_{\parallel}/T_{\perp}}$.
The dichroism was estimated for the Amide bands: for the Amide A  $\Delta"\simeq 0.014$ suggests only a minute transmission
change $\sqrt{T_{\parallel}/T_{\perp}}\simeq 98.6\%$ for the two perpendicular polarisations. Similarly, the results obtained for the Amide I ($\Delta"\simeq 0.027$ and 97.4\%) and Amide II ($\Delta"\simeq 0.019$ and 98.1\%) also indicated that very small changes of absorbance of light passed through the thin $100$~nm lateral slices of the silk fiber. This shows that anisotropy of absorbance can be measured from nanoscale materials of sub-wavelength thickness. \blue{There were no apparent spectral differences among few measured samples at different orientations of 100-nm-thick silk slices. The far-field (Fig.~\ref{f-100}) and near-field (Fig.~\ref{f-spectra}) absorbance spectra are comparable and are matching earlier results measured from thicker samples~\cite{17mre115028}. This study shows that the SNOM measurements reach resolution required to measure structural composition of silk fibres which correspond to the crystalline segments observed in XRD and the measurement can be carried out from nano-thin slices of silk. }

\section{Conclusions and outlook}

In summary, spectral characterisation, lateral mapping and
transmission with deep sub-wavelength resolution at IR molecular
fingerprinting spectral window were demonstrated using thin $100$~nm
lateral slices of silk. Absorbance and reflectance spectra of silk
with resolution of SNOM tip $\sim 10$~nm were obtained. Absorbance
from nano-thin silk slices with only 1.5\% of the wavelength were
measured when the beam diameter was comparable with the
IR-wavelength. Hyper-spectral mapping across the silk fiber slice
was obtained at a high accuracy and reproducibility. Orientational map of the amide bands was revealed \blue{and was consistent with data collected from bulk samples. It shows that preparation of thin micro-tome slices of soft bio-materials is not altering their structure and opens possibility to read optical properties from nano-volumes. In the case of optical measurements, optical averaging over thicker inhomogeneous volumes of samples can be avoided using nano-slices and this provides more reliable direct measurement of optical properties}. The study demonstrated 
characterisation of the silk fiber with a nano-scale resolution in all three dimensions.



\begin{acknowledgements}
\small{JM acknowledges a partial support by a JSPS KAKENHI Grant
No.16K06768 and 18H04506. We acknowledge partial support via ARC Discovery
DP170100131 grant. Experiments were carried out through a beamtime
proposal (ID. 12107) at the Australian Synchrotron IRM Beamline, part of ANSTO. We are grateful for R.~Kikuchi from Materials
Analysis Division of Tokyo Institute of Technology, Ookayama, for his assistance with ultramicrotomy. X-ray characterisation was performed at the Stanford Nano Shared Facilities (SNSF), supported by the National Science Foundation
under award ECCS-1542152. SJ is grateful for sabbatical stays at
Tokyo Institute of Technology and Shizuoka University. A part of this work was carried out under the Cooperative Research
Project Program of the Research Institute of Electronics, Shizuoka University. 
}
\end{acknowledgements}

\bibliographystyle{bjnano}
\small
\bibliography{nanofab1,nan,paper6b}
\end{document}